\documentclass[12pt]{article}

\usepackage{authblk}
\usepackage{slashed}
\usepackage{amsmath}
\usepackage{amssymb}

 \newcommand{\texteq}[1]{\mbox{$\;#1\;$}}
 \newcommand{\bigket}[1]{\bigl| #1\bigr\rangle}

\title{The Schwinger Model in Point Form}
\author[1]{D. Kupelwieser}
\author[1]{W. Schweiger}
\author[2]{W.H. Klink}
\affil[1]{\small{Institut f\"ur Physik, Universit\"at Graz, A-8010 Graz, Austria}}
\affil[2]{\small{Dept. Physics and  Astronomy, The University of Iowa,  Iowa City, IA 52242-1479, U.S.}}
\date{}

\begin{document}

\maketitle

\setlength{\parindent}{0pt}
\setlength{\parskip}{1ex}

\begin{abstract}
We attempt to solve the Schwinger model, i.e. massless QED in 1+1
dimensions, by quantizing it on a space-time hyperboloid $x_\mu
x^\mu = \tau^2$. The Fock-space representation of the 2-momentum
operator is derived and its algebraic structure is analyzed. We
briefly outline a solution strategy.
\end{abstract}

\section{Introduction}

The Schwinger model is quantum electrodynamics of massless
fermions in 1  space and 1 time dimension~\cite{schwinger} and
serves as a popular testing ground for non-perturbative methods in
quantum field theory (QFT). It is an exactly solvable,
super-renormalizable gauge theory that exhibits various
interesting phenomena~\cite{cjs}, such as confinement, which one
would like to understand better in 1+3-dimensional QFTs.
Originally it was solved by means of functional
methods~\cite{schwinger}. Later on also operator solutions were
found~\cite{lowensteinswieca} and spectrum and eigenstates of the
theory were calculated by quantizing it at equal time
$x^0=\mathrm{const.}$~\cite{manton,Link:1990bc} or at equal
light-cone time $x^+=x^0+x^1=\hbox{const.}$~\cite{Lenz:1991sa}. We
rather attempt to solve the Schwinger model by means of canonical
quantization on the space-time hyperboloid $x_0^2-x_1^2=\tau^2$.
Each of these quantization hypersurfaces is associated with a
particular form of relativistic Hamiltonian dynamics~\cite{dirac},
namely the instant form, the front form and the point form,
respectively.

The quantization surface in point form is a space-time hyperboloid
which  is invariant under the action of the Lorentz group. The
kinematic (interaction independent) generators of the Poincar\'e
group are therefore those of the Lorentz subgroup. All the
interactions go into the components of the 2-momentum $P^\mu$,
i.e. the generators of space-time translations, which provide the
dynamics of the system. One of the main virtues of point-form
dynamics is obviously a simple behavior of wave functions and
operators under Lorentz transformations. This has already been
exploited in applications to relativistic few-body
systems~\cite{bks}, but corresponding studies of interacting
quantum field theories are still very sparse. The best-known
paper is that of Fubini et al.~\cite{Fubini:1972mf}, who deal with
point-form QFT in 2-dimensional Euclidean space-time. We rather
want to extend equal-$\tau$ quantization in Minkowski space-time,
as it was worked out in Ref.~\cite{biernat} for free field
theories, to the interacting case. The solution being known, the
Schwinger model would be an interesting example to test the
point-form approach against other methods. The hope is then that
point-form quantum field theory will eventually represent a useful
alternative in the study of 4-dimensional quantum field theories.

The Lagrangian of the Schwinger model is
\begin{equation}\label{lagrangian}
\mathcal{L} =
\mathcal{L}_{\gamma}+\mathcal{L}_{e}+\mathcal{L}_{\mathrm{int}}
=\underbrace{-\,\frac{1}{4}F^{\mu\nu}F_{\mu\nu}}_{\text{photon
part}} + \underbrace{\frac{i}{2}\,\bar{\psi}\,
\stackrel{\leftrightarrow}{\slashed{\partial}}\,
\psi}_{\text{fermion part}} +
\underbrace{\frac{1}{2}\,e\,\bar{\psi}\,
\slashed{A}\,\psi}_{\text{interaction part}}
\end{equation}
with the $2\times 2$ Dirac matrices being represented, as usual,
in the  Weyl basis, i.e.  \texteq{\gamma^0=\sigma_1},
\texteq{\gamma^1=i\sigma_2} and
\texteq{\gamma^5=\gamma^0\gamma^1=-\sigma_3}.

\section{The 2-Momentum Operator}

\subsection{The free part}

This exposition follows closely Ref.~\cite{biernat} to which we
refer for  further details.

\paragraph{Fermions:}

In order to obtain the Fock-space representation of the free
fermion 2-momentum  operator, we Fourier-expand the Dirac field
$\psi(x)$ in terms of plane waves using  the fermion and
antifermion annihilation (creation) operators $c^{(\dagger)}(p)$
and $d^{(\dagger)}(p)$ and the spinor basis $\{u(p), v(p)\}$. In
the massless case, the spinors are ($p^0=|p^1|$):
\begin{equation}
 u(p)=\frac{1}{\sqrt{2p^0}}\binom{p^0-p^1}{p^0+p^1}
\qquad\text{and}\qquad
v(p)=\frac{1}{\sqrt{2p^0}}\binom{p^1-p^0}{p^1+p^0}\;.
\end{equation}
The free fermion  2-momentum operator in point-form is then
obtained from  the stress-energy tensor $\Theta^{\mu\nu}_{e}$ by
integrating over the space-time hyperboloid $x_\mu x^\mu =
\tau^2$:
\begin{equation}\label{2mom}
P^\mu_e =  \int_{\mathbb{R}^2} \underbrace{2 d^2x\
\delta\left(x^2-\tau^2\right)\ \theta(x^0)\
x_\nu}_{\text{point-form \lq\lq  surface\rq\rq\ element}}\
\Theta^{\nu\mu}_{e}\, , \quad \hbox{with}\quad \Theta^{\nu\mu}_e =
\frac{i}{2}\,\bar\psi\,\gamma^\nu\,\stackrel{\leftrightarrow}{\partial}
\!\!\!\phantom{l}^\mu\,\psi\, .
\end{equation}
Inserting now the plain-wave expansion for the fields and
interchanging  momentum and $x$ integrations we are left with the
covariant distribution
\begin{eqnarray}\label{Wmu}
W_\nu(q) &=&
2 \int_{\mathbb{R}^2} d^2x\ \delta(x^2-\tau^2)\ \theta(x^0)\
x_\nu\ e^{-iqx}\nonumber \\
&=& 2 \pi \delta(q^2) \epsilon(q^0) q_\nu + 2 \pi \theta(q^2)
\delta(q^0) J_0(\tau \sqrt{q^2}) g_{\nu 0}\\ & &\hspace{-0.3cm} - \frac{\pi
\tau}{\sqrt{q^2}}\theta(q^2)\left[i Y_1(\tau \sqrt{q^2})+
\epsilon(q^0) J_1(\tau \sqrt{q^2}) \right] q_\nu - \frac{2 i
\tau}{\sqrt{-q^2}}\theta(-q^2) K_1(\tau \sqrt{-q^2})\, q_\nu\, .\nonumber
\end{eqnarray}
When evaluating equation \eqref{2mom} for the free parts of the
Lagrangian \eqref{lagrangian}, $W_\nu$ is contracted with
spinor products of the form $\bar{u} \gamma^\nu u$, $\bar{u} \gamma^\nu v$, etc. All the contractions with $q_\nu$ vanish and only the term $\propto\ \theta(q^2)\delta(q^0)\,g_{\nu
0}$ survives. The result, as already shown by Biernat et al.
\cite{biernat} using a different trick to evaluate $W_\nu$, is
(after normal ordering)
\begin{equation}
P^\mu_e = {\int\frac{dp^1}{2p^0}}\:p^\mu\,\Bigl(c^\dagger(p)\,c(p)
+ d^\dagger(p)\,d(p)\Bigr)\, ,
\end{equation}
i.e. the same as in instant form.

\paragraph{Photons:}
For the free photon 2-momentum operator  we proceed in an
analogous way.~\footnote{See also Ref.~\cite{murphy} for a detailed derivation of the gluon 2-momentum operator.}
The Fourier expansion of the vector potential
$A^\mu(x)$ in terms of plane waves gives rise to the photon
creation- and annihilation operators $a_\kappa^\dagger(k)$ and
$a_\kappa(k)$ and to polarization vectors
$\epsilon^\mu_\kappa(k)$, with $\kappa=0,1$ labeling the
polarization. The polarization vectors are orthonormalized
according to $\epsilon^\mu_{\kappa^\prime}(k)
\epsilon_{\kappa\mu}(k)=g_{\kappa^\prime \kappa}$. In order to
preserve the nice covariance properties of the point form, we work
within the Lorenz gauge and use the Gupta-Bleuler quantization
procedure. As a consequence there are no physical photons left.
The 0- and the 1-component of the photon field are pure gauge degrees of freedom.
Proceeding in analogy to the fermion part we find for
the Fock-space representation of the free photon 2-momentum
operator again the same result as for equal-time quantization,
i.e.
\begin{equation}
P^\mu_\gamma = \sum_{\kappa=0}^1\int\frac{dk^1}{2k^0}\,k^\mu\,g^{\kappa\kappa}
a^\dagger_\kappa(k)\,a_\kappa(k)\, .
\end{equation}

\subsection{The interaction part}

Since there is no derivative in the interaction part of the
Lagrangian \eqref{lagrangian}, the interaction part of the
stress-energy tensor is simply given by
\texteq{\Theta_{\text{int}}^{\mu\nu}=-g^{\mu\nu}\,\mathcal{L}_{\text{int}}}.
The interaction part of the 2-momentum operator is then
\begin{equation}\label{Pmuint}
P^\mu_{\text{int}}=-\int_{\mathbb{R}^2}2\,d^2x\,\delta(x^2-\tau^2)\,
\theta(x^0) \,x^\mu\,\mathcal{L}_{\text{int}}(x)\, .
\end{equation}
One can check explicitly that the corresponding integral for the
interaction part of the boost generator vanishes as
expected~\cite{biernat}.

To obtain the Fock-space representation of $P^\mu_{\text{int}}$ we
proceed as before. The only difference is now that $W_\nu(q)$ does
not provide a momentum conserving $\delta$ function. But this is
not surprising. Both components of the momentum operator are
interaction dependent so that one cannot expect momentum
conservation at interaction vertices. But what one can do is to
analyze the algebraic structure of $P^\mu_{\text{int}}$. By
appropriately collecting terms it can be cast into the form
\begin{equation}\label{Pmuintres}
 P^\mu_{\mathrm{int}} = -e\sum_{\kappa=0}^1\int \frac{dk^1}{2 k^0}
\left(\mathcal{A}(X_{\kappa}^\mu)(k)\,a_{\kappa}(k) +
\mathcal{A}^\dagger(X_{\kappa}^\mu)(k)\,a^\dagger_{\kappa}(k)\right)
\end{equation}
with
\begin{equation}
\mathcal{A}(X_{\kappa}^\mu)(k)=\int \frac{dp^1}{2p^0}\int \frac{dp^{\prime 1}}{2p^{\prime 0}}\, \left(
c^\dag (p'), d(p) \right) X^\mu_{(\kappa)}(k,p',p) \left(
\begin{array}{l}c(p)\\ d^\dag(p)\end{array}\right)
\end{equation}
The distribution $W^\mu$ for different combinations of the momenta
$p$, $p^\prime$ and $k$ together with the different spinor
products determines essentially the elements of the $2\times2$
matrix $X^\mu_{(\kappa)}(k,p',p)$.

\section{The Eigenvalue Problem}
%
Putting all the pieces together we finally end up with the
eigenvalue problem
\begin{eqnarray}
\hspace{-1cm}\left(P^\mu_e+P^\mu_\gamma+P^\mu_{\mathrm{int}}\right)\bigket{\Psi} &=&
\mathcal{A}(E^\mu)\bigket{\Psi}+\sum_{\kappa=0}^1\int
\frac{dk^1}{2 k^0} \Big(k^\mu g^{\kappa\kappa}
a^\dagger_\kappa(k)\,a_\kappa(k)\nonumber\\ &&- e
\mathcal{A}(X_{\kappa}^\mu)(k)\,a_{\kappa}(k) -e
\mathcal{A}^\dagger(X_{\kappa}^\mu)(k)\,
a^\dagger_{\kappa}(k)\Big)\bigket{\Psi} =p^\mu\bigket{\Psi}
\end{eqnarray}
which we want to solve non-perturbatively. Here we have also
expressed the fermion kinetic energy in terms of the
$\mathcal{A}$s to emphasize that the fermion creation and
annihilation operators occur only in bilinear combinations. The
argument $E^\mu$ is essentially a diagonal matrix containing $\pm
\delta(p^{1\prime}-p^1)$.

A possible strategy to solve this eigenvalue problem was proposed
in Ref.~\cite{klink}. The first step is to keep the number of
modes finite. This could, e.g., be done without spoiling
Lorentz-transformation properties by compactifying the $x^1$
direction such that one ends up with a deSitter space. A finite
number of modes means also that only a finite number of
fermion-antifermion pairs can be created. In order to keep the
number of bosons finite the boson algebra is then considered as a
contraction limit of another unitary algebra that restricts the
number of bosons in any mode. In this way one ends up with a
solvable algebraic problem that involves only a finite number of
modes and a finite number of particles. The interesting question
will be whether the well known results for Schwinger model are
recovered upon performing the necessary contractions that restore
the original theory.

\paragraph{Acknowledgement:} D. Kupelwieser acknowledges the support of the \lq\lq Fonds zur F\"order\-ung der wissenschaftlichen Forschung in \"Osterreich\rq\rq\ (FWF DK W1203-N16).


\end{document}